\documentclass[journal=jacsat,manuscript=article]{achemso} 
\usepackage[version=3]{mhchem} 
\setkeys{acs}{articletitle = true}

\usepackage{graphicx}%
\usepackage{multirow}%
\usepackage{amsmath,amssymb,amsfonts}%
\usepackage{amsthm}%
\usepackage{mathrsfs}%
\usepackage[title]{appendix}%
\usepackage{xcolor}%
\usepackage{textcomp}%
\usepackage{manyfoot}%
\usepackage{booktabs}%
\usepackage{algorithm}%
\usepackage{algorithmicx}%
\usepackage{algpseudocode}%
\usepackage{listings}%
\usepackage{atbegshi}
\addtocounter{page}{-1}
\AtBeginDocument{\AtBeginShipoutNext{\AtBeginShipoutDiscard}}

\usepackage{mathtools}
\usepackage{upgreek}
\usepackage{thmtools}
\usepackage{lineno}
\usepackage{dsfont}
\usepackage{hyperref}
\usepackage[normalem]{ulem}
\usepackage{setspace}
\usepackage{outline}
\usepackage{physics}

\usepackage{siunitx}
\usepackage{tikz}
\usetikzlibrary{shapes.geometric, arrows,positioning}
\usepackage{qcircuit}
\usepackage{lipsum}
\usepackage{chemfig}
\usepackage{silence}
\usepackage{algorithm}
\usepackage{algpseudocode}

\usepackage{lineno}

\newcommand{\pool}{\mathcal{P}}

\def\app#1#2{%
  \mathrel{%
    \setbox0=\hbox{$#1\sim$}%
    \setbox2=\hbox{%
      \rlap{\hbox{$#1\propto$}}%
      \lower1.1\ht0\box0%
    }%
    \raise0.25\ht2\box2%
  }%
}

\begin{tocentry}
	\includegraphics[width=\textwidth]{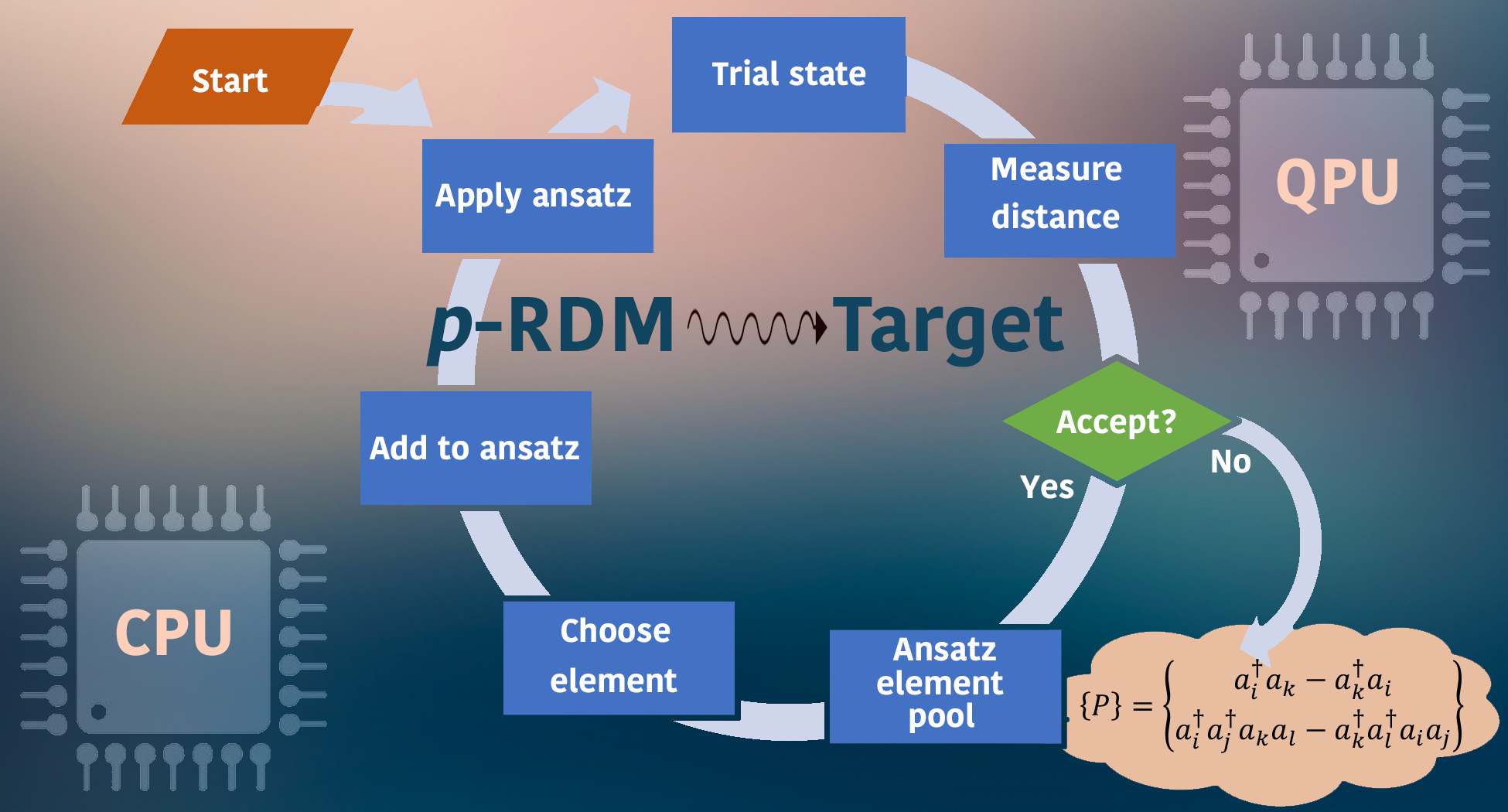}
\end{tocentry}

\title[Article Title]{Determining the $N$-representability of a reduced density matrix via unitary evolution and stochastic sampling}
\author{Gustavo E. Massaccesi}
\affiliation{Departamento de Ciencias Exactas, Ciclo B\'asico Com\'un, Universidad de Buenos Aires, Ciudad Universitaria, 1428 Buenos Aires, Argentina}
\alsoaffiliation{Instituto de Investigaciones Matem\'aticas ``Luis A. Santal\'o'' (IMAS), Consejo Nacional de Investigaciones Cient\'ificas y T\'ecnicas, Universidad de Buenos Aires. Ciudad Universitaria, 1428 Buenos Aires, Argentina}

\author{Ofelia B. O\~{n}a}
\affiliation{Instituto de Investigaciones Fisicoqu\'imicas Te\'oricas y Aplicadas, Universidad Nacional de La Plata, Consejo Nacional de Investigaciones Cient\'ificas y T\'ecnicas. Diag. 113 y 64 (S/N), Sucursal 4, CC 16, 1900 La Plata, Argentina}

\author{Pablo Capuzzi}
\affiliation{Universidad de Buenos Aires, Facultad de Ciencias Exactas y Naturales, Departamento de F\'isica. Ciudad Universitaria, 1428 Buenos Aires, Argentina}
\alsoaffiliation{CONICET - Universidad de Buenos Aires, Instituto de F\'isica de Buenos Aires (IFIBA). Ciudad Universitaria, 1428 Buenos Aires, Argentina}

\author{Juan I. Melo}
\affiliation{Universidad de Buenos Aires, Facultad de Ciencias Exactas y Naturales, Departamento de F\'isica. Ciudad Universitaria, 1428 Buenos Aires, Argentina}
\alsoaffiliation{CONICET - Universidad de Buenos Aires, Instituto de F\'isica de Buenos Aires (IFIBA). Ciudad Universitaria, 1428 Buenos Aires, Argentina}

\author{Luis Lain}
\affiliation{Departamento de Qu\'imica F\'isica, Facultad de Ciencia y Tecnolog\'ia, Universidad del Pa\'is Vasco. Apdo. 644, E-48080 Bilbao, Spain}

\author{Alicia Torre}
\affiliation{Departamento de Qu\'imica F\'isica, Facultad de Ciencia y Tecnolog\'ia, Universidad del Pa\'is Vasco, Apdo. 644, E-48080 Bilbao, Spain}

\author{Juan E. Peralta}
\email{juan.peralta@cmich.edu}
\affiliation{Department of Physics, Central Michigan University, Mount Pleasant, MI, 48859, USA}

\author{Diego R. Alcoba}
\email{dalcoba@df.uba.ar}
\affiliation{Universidad de Buenos Aires, Facultad de Ciencias Exactas y Naturales, Departamento de F\'isica. Ciudad Universitaria, 1428 Buenos Aires, Argentina}
\alsoaffiliation{CONICET - Universidad de Buenos Aires, Instituto de F\'isica de Buenos Aires (IFIBA). Ciudad Universitaria, 1428 Buenos Aires, Argentina}

\author{Gustavo E. Scuseria}
\email{guscus@rice.edu}
\affiliation{Department of Chemistry, Rice University, Houston, TX 77005-1892}
\alsoaffiliation{Department of Physics and Astronomy, Rice University, Houston, TX 77005-1892}


\AtBeginDocument{\RenewCommandCopy\qty\SI}

\begin{document}

\begin{abstract}
The $N$-representability problem consists in determining whether, for a given $p$-body matrix, there exists at least one $N$-body density matrix from which the $p$-body matrix can be obtained by contraction, that is, if the given matrix is a  $p$-body reduced density matrix ($p$-RDM).
The knowledge of all necessary and sufficient conditions for a $p$-body matrix to be $N$-representable allows the constrained minimization of a many-body Hamiltonian expectation value with respect to the $p$-body density matrix and, thus, the determination of its exact ground state.
However, the number of constraints that complete the $N$-representability conditions grows exponentially with system size, and hence the procedure quickly becomes intractable for practical applications.
This work introduces a hybrid quantum-stochastic algorithm to  effectively 
replace the $N$-representability conditions. 
The algorithm consists of applying to an initial $N$-body density matrix a sequence of unitary evolution operators
constructed from a stochastic process that successively approaches the reduced state of the density matrix on a $p$-body subsystem, represented by a $p$-RDM, to a target $p$-body matrix, potentially a $p$-RDM.
The generators of the evolution operators follow the well-known adaptive derivative-assembled pseudo-Trotter method (ADAPT), 
while the stochastic component is implemented using a simulated annealing process. 
The resulting algorithm is independent of any underlying Hamiltonian, and  
it can be used to decide if a given $p$-body matrix is $N$-representable, establishing a criterion to determine its quality and correcting it. 
We apply the proposed hybrid ADAPT algorithm to alleged reduced density matrices  from a quantum chemistry electronic Hamiltonian, from the reduced Bardeen–Cooper–Schrieffer model with constant pairing, and from the Heisenberg XXZ spin model.  
In all cases, the proposed method behaves as expected for 1-RDMs and 2-RDMs,
evolving the initial matrices towards different targets.

\end{abstract}

\newpage
\section{Introduction}\label{sec1}

The $N$-representability problem has been identified as one of the most important challenges in electronic structure theory. \cite{Stillinger.book}
 Since only the 2-body reduced density matrix (2-RDM)  needs to be determined to calculate the energy of a pairwise interacting system (such as interacting electrons in atoms, molecules, or condensed matter)
 \cite{Coleman.book.2000,Mazziotti.book.2007}, it is common to focus the $N$-representability problem on 
 finding the necessary and sufficient constraints or conditions for the 2-RDM to ensure that it arises from the reduction of an $N$-body density matrix
 \cite{Coleman.RevModPhys.1963}.
These conditions are known as the $N$-representability conditions
\cite{Mazziotti.book.2007,Coleman.book.2000,Coleman.RevModPhys.1963,Garrod.JMP.1963,Kummer.JMP.1967,Erdahl.IJQC.1978,Erdahl.book.1987,Mazziotti.PRL.2012}, 
and a failure to satisfy them causes a collapse of variational approaches, potentially leading to energies below their true ground state. 
The $N$-representability problem can be categorized as belonging to the quantum Merlin-Arthur complete class, a generalization of the (nondeterministic) polynomial-time complete class.\cite{Liu.PRL.2007} 
This emphasizes its poor computational tractability for practical calculations. Recent advances in this field 
\cite{Mazziotti.PRL.2012,Mazziotti.PRA.2016,Mazziotti.2023}
have enabled more accurate approximations that utilize $N$-representability conditions able to target strongly correlated systems. However, the problem's intrinsic complexity makes this approach only accessible to a relatively small set of systems. 
In general, the $N$-representability problem is presented for a $p$-RDM derived from contracting $(N-p)$ degrees of freedom of an $N$-body pure state or density matrix (single wave function) or of an $N$-body ensemble (convex linear combination) of pure states or density matrices.
\cite{Coleman.book.2000}
Although most works have been historically focused on ensemble $N$-representability, some recent papers tackle the pure $N$-representability problem.{\cite{Altunbulak.CMP.2008,Mazziotti.PRA.2016,Mazziotti.book.2007,Coleman.IJQC.1978,Schilling.2013,Chakraborty.2014,Theophilou.2015,Schilling.2018,Smart.2019,Boyn.2019,Avdic.2023}

The adaptive derivative-assembled pseudo-Trotter (ADAPT) variational quantum eigensolver (VQE)
has been introduced to enable electronic structure quantum simulations in near-term quantum hardware.\cite{Grimsley.NatComm.2019}
Starting from an initial wave function ansatz, the algorithm evolves the ansatz by successively applying unitary transformations depending on 1- and 2-body operators selected from a predefined pool, while a variational parameter that minimizes the total energy is introduced at every step.  One of the advantages of the ADAPT-VQE is that the number of resulting variational parameters is small, which makes its implementation in quantum computers attainable with shallow-depth quantum circuits.\cite{Grimsley.NatComm.2019,Fedorov.2022,Bharti.2022}
Other approaches related to} the ADAPT-VQE including the contracted quantum eigensolvers (CQE) \cite{Smart.2021,Warren.2024}
have been successfully applied to several problems ranging from quantum chemical simulations, strongly correlated systems, and wave function optimization.\cite{Grimsley.NatComm.2019,Yordanov.2021,Tang.2021,Liu.2022,Stein.2022,Ziems.2024,Vaquero.2024}

In this work, we introduce a hybrid ADAPT variational quantum algorithm (VQA)  that enables an initial $p$-RDM to evolve  towards a target $p$-body matrix (alleged $p$-RDM) and approach it as much as possible, using the Hilbert-Schmidt distance as a measure.
Other hybrid strategies which directly determine (approximately) N-representable RDMs on a quantum device have also been proposed in the literature\cite{Smart.pra.2019,Smart.2021,Smart.pra.2022}. Our proposed algorithm incorporates elements from the ADAPT methodology to properly evolve the $p$-RDM and relies on stochastic optimization to minimize the distance to avoid barren plateaus in the search process. 
Importantly, the proposed ADAPT-VQA is independent of any underlying Hamiltonian, can be used to determine the quality of an alleged $p$-RDM and to correct it, and is robust under statistical noise. 


\section{Theory and Algorithm}\label{sec2}

\label{sec:VQENRepSection}
The underlying idea of our VQA is to utilize ansatz circuits, which are characterized by a collection of vector parameters, $\{\vec{\theta}\}$, for producing entangled trial $N$-body (pure) states or density matrices, denoted as $\rho(\{\vec{\theta}\})$. A classical stochastic global search algorithm is employed to adjust the vector parameters $\{\vec{\theta}\}$ and minimize the cost function
given by the Hilbert-Schmidt distance $D$ between the (physical) reduced state of  $\rho(\{\vec{\theta}\})$ on a $p$-body subsystem, represented by a $p$-body reduced density matrix $^p\rho(\{\vec{\theta}\})\,\equiv\,p!\binom{N}{p}\Tr_{N-p}[\rho(\{\vec{\theta}\})]
$, and a fixed target (possibly physical) represented by   
$^p\rho_t$.
Provided that the ansatz is sufficiently expressive
so that it contains a circuit that well-approximates the optimal solution, this distance satisfies
\begin{equation}
    \label{eq:RayleighRitz}   
D(^p\rho(\{\vec{\theta}\}),\,^p\rho_t)
=\Tr[(^p\rho(\{\vec{\theta}\})-\,^p\rho_t)^2] =  \|^p\rho(\{\vec{\theta}\})-\,^p\rho_t\|_2^2\geq 
\mathcal{D}_{0},
\end{equation}
which allows $\min_{\{\vec{\theta}\}} (D(^p\rho(\{\vec{\theta}\}),\,^p\rho_t))$ to approach the minimal feasible distance $\mathcal{D}_{0}$.
Importantly,  $\mathcal{D}_{0}=0$ if and only if $^p\rho_t$ is pure $N$-representable, and hence it can be used as a measure of the degree of $N$-representability of $^p\rho_t$. 
Moreover, the evolved $\rho(\{\vec{\theta}\})$ can be used to obtain the $N$-representable (physical) reduced state $^p\rho$ closest to $^{p}\rho_{t}$. 
Since $^{p}\rho_{t}$ is not necessarily $N$-representable, this highlights a potential use of the algorithm: to correct a non-$N$-representable matrix.

To perform the minimization of the distance $D$, we take a hybrid approach.
A classical stochastic optimizer iteratively constructs the ansatz generating the parameterized state, while a quantum computer calculates the Hilbert-Schmidt-distance value for this parameterized state.
At each iteration step $n$, 
the state generated by a set of parameters $\{\vec{\theta}\}_n$ can be written as 
\begin{equation}
  \label{eq:TrialStates}
\rho_{n}(\{\vec{\theta}\}_n)=
  \rho_{n}(\vec{\theta}_1,\ldots,\vec{\theta}_n) = 
U_{n}(\vec{\theta}_1,\ldots,\vec{\theta}_n)\rho_{0}\,U_{n}^{\dagger}(\vec{\theta}_1,\ldots,
\vec{\theta}_n)\,,
\end{equation}
where $\rho_0$ is an initial state and    $U_{n}$ represents a unitary transformation that depends on the parameters $\{\vec{\theta}\}_n$, which results of the successive actions of single unitary transformations. 
Eq.~(\ref{eq:TrialStates}) evidences that since $\rho_n(\{\vec{\theta}\}_n)$
 is parameterized by $n$ vector parameters, then  the corresponding $p$-body reduced state  is also parameterized by the same $n$ vector parameters, $^p\rho_{n}(\{\vec{\theta}\}_n)$. 
In our ADAPT-VQA, 
a stochastic optimizer,  implemented on a classical computer, and 
a quantum algorithm, realized on a quantum computer, operate in conjunction to find a minimum-distance 
value of $D$:
\begin{equation}
  \label{eq:EnergyExpectation}
D_n \equiv \min_{\{\vec{\theta}\}_n}
\Tr[(^p\rho_n(\{\vec{\theta}\}_n)-\,^p\rho_t)^2] =  \min_{\{\vec{\theta}\}_n}\|^p\rho_n(\{\vec{\theta}\}_n)-\,^p\rho_t\|_2^2
.
\end{equation}
In other words, the ADAPT-VQA gradually incorporates parametrized elements into its ansatz to form $^p\rho_{n}(\{\vec{\theta}\}_n)$ such that $D_n$ approaches  $\mathcal{D}_0$ as $n \rightarrow \infty$.

\begin{figure*}[h]
\includegraphics[width=0.9\linewidth]{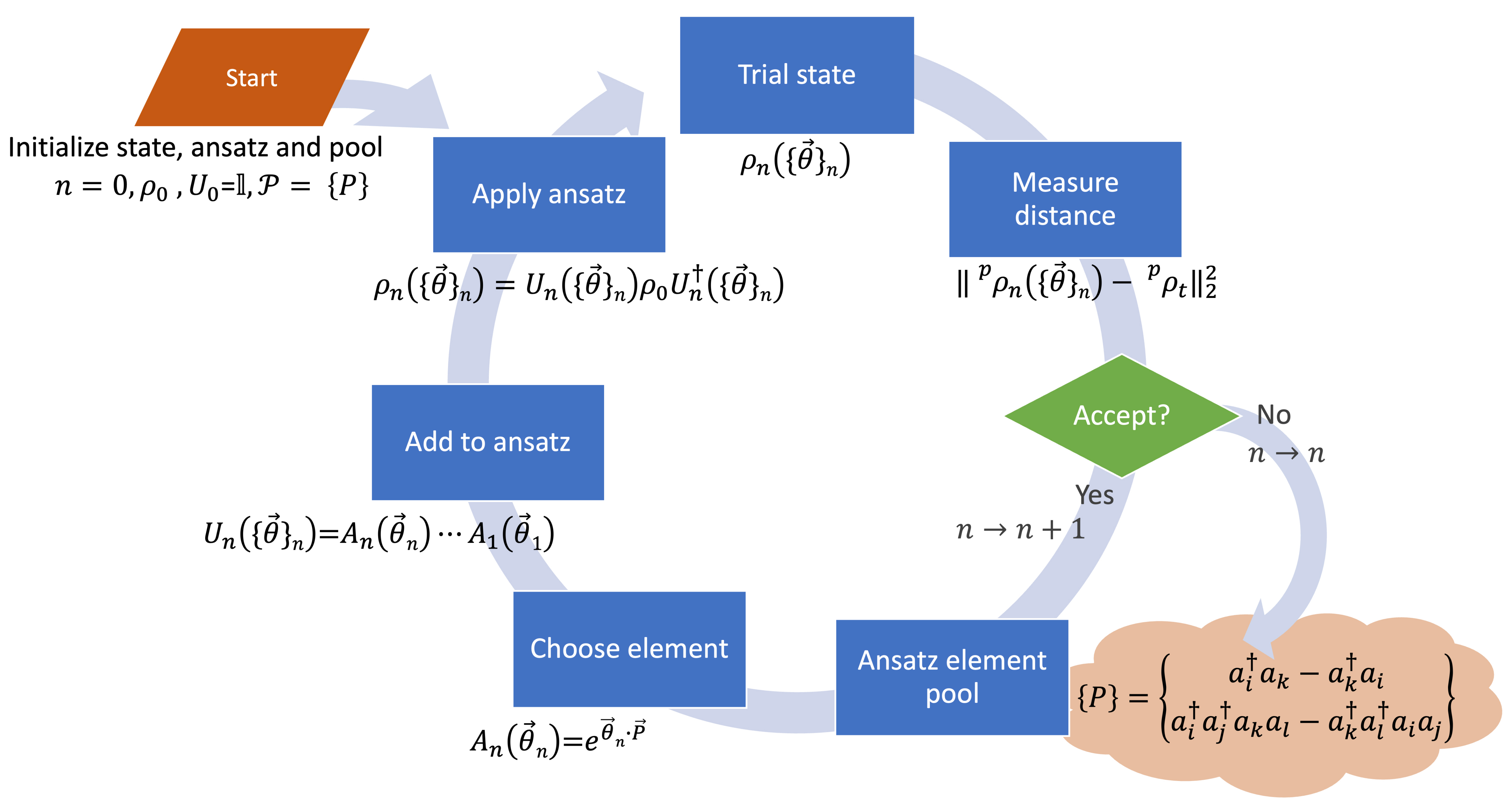}
\caption{ADAPT-VQA flowchart. 
\label{flochart}}
\end{figure*}

The iterative ansatz is constructed by first  initializing  the ADAPT-VQA to  a state $\rho_{0}$, usually an independent-particle-model state, and then generating  a series of trial states by  adding, one at the time,  elements of the form
\begin{linenomath} \begin{align}
    \label{eq:AnsatzElement}
    A_{\alpha}(\vec\theta_{\alpha}) = e^{\vec{\theta}_{\alpha} \cdot \vec{P}},
\end{align} \end{linenomath}
where  $\vec{P}$ is a $\pool$-dimensional vector
embedding a finite pool $\pool$ of
anti-hermitian operators (see next Section), and $\vec\theta_{\alpha}$ is a $\pool$-dimensional vector parameter that has all zero elements except one randomly chosen with a random amplitude in the interval  $[-\theta_{max},\theta_{max}]$, with $\theta_{max}$ 
decreasing or increasing as the iteration number increases depending on the local features of the cost function.
Thus, the unitary ansatz evolves using the simple prescription
\begin{linenomath} \begin{align}
    U_0 &= \mathds{I} , \\
    U_n(\{\vec{\theta}\}_n) &= 
A_n(\vec\theta_n)U_{n-1}(\{\vec{\theta}\}_{n-1}).
\end{align} \end{linenomath}
At each iteration, the resulting 
unitary ansatz is accepted with a certain probability, which is initially high and gradually decreases as the number of iterations increases. This process ensures that 
the Hilbert-Schmidt distance $D_n$ progressively decreases and  potentially avoids barren plateaus found in gradient-descent methods.\cite{McClean.NC.2018}  
The distance $D_n$  is evaluated in a quantum computer after adding $A_{n}(\vec\theta_n)$ in the $n-th$ step as
\begin{equation}
  \label{eq:LocalEnergyExpectation}
D_{n}
(\vec{\theta}_n)
\equiv 
\Tr[(\,p!\binom{N}{p}\Tr_{N-p}[A_{n}(\vec\theta_n) \rho_{n-1} 
(\{\vec{\theta}\}_{n-1})
A_{n}^{\dagger}(\vec\theta_n)]\,-\,^p\rho_t)^2] \,.
\end{equation}
Provided that the difference $D_n - D_{n-1}$ is greater than a distance precision $\epsilon$, the iterative algorithm proceeds. If $D_n - D_{n-1}$ is less than or equal to $\epsilon$ for a certain number of consecutive steps, the algorithm terminates at a final length $n=L$ and produces $D_L \equiv 
D_n$ as the numerical estimate of $\mathcal{D}_0$. The algorithm flowchart is provided in Fig.~\ref{flochart}.

\section{Simulations}\label{sec22}

To explore the applicability of our ADAPT-VQA, we have chosen three representative strongly correlated systems typically used in quantum chemistry, nuclear structure, and condensed matter physics:\cite{Rubio-Garcia.JCTC.2018,Rubio-Garcia.JCP.2019,Alcoba.2024,Alcoba.jcpa.2024} the linear H$_4$ molecule, the reduced Bardeen–Cooper–Schrieffer (BCS) or constant pairing model, and the Heisenberg XXZ spin model, respectively. 
With this choice, we also test the capability of our ADAPT-VQA with systems that are not well described by unitary coupled cluster theories.\cite{Harsha.2018, Lee.jctc.2019}

In this work, we focus on a fermionic ADAPT-VQA where the pool of excitation operators can be reduced to 
{\cite{Mazziotti.pra.2007,OMalley.PRX.2016,Barkoutsos.PRA.2018,Evangelista.JCP.2019}
\begin{align}
  \label{eq:fermionic}
  P_{i k} &= a_{i}^{\dagger}a_{k}-a_{k}^{\dagger} a_{i}, \; \; 
\end{align}
and
\begin{align}
  \label{eq:fermionic2}
  P_{i j k l} &= a_{i}^{\dagger} a_{j}^{\dagger} a_{k} a_{l}
                -a_{k}^{\dagger} a_{l}^{\dagger} a_{i} a_{j}\,,
\end{align}
Here, $a_{i}^{\dagger}$ and $ a_{i}$ are fermion creation and annihilation operators acting on
an orthonormal finite single-particle basis $\{i,j,k,l,...\}$, represented in terms of Pauli operators using the Jordan-Wigner transformation as 
is paramount to simulate fermionic systems on quantum devices.\cite{Tilly.2022}

{\it Computational details:} 
All the  computations in this work were carried out simulating an ideal, noiseless quantum device
using an in-house code that interfaces  {\sc PySCF}
\cite{pyscf_wires,pyscf_jcp}
and the OpenFermion module of {\sc PySCF}  for integral computations and the OpenFermion 
code 
\cite{OpenFermion2020}
for the  Jordan–Wigner mappings.
The pool of operators consists of spin-adapted generalized singles and doubles, with double excitation operators decomposed into singlets and triplets.\cite{Grimsley.NatComm.2019,Ramoa.2022}
This choice, which is due to the nature of the treated systems, reduces significantly the number of operators in the pool for hard-core bosons and spin
models based on the $SU(2)$ pairing algebra.
All  calculations involving electrons were performed using the STO-3G basis set
\cite{sto3g}.

The stochastic optimization uses the simulated annealing (SA) algorithm 
introduced by Kirkpatrick {\it et al.}
\cite{Kirkpatrick}.  
The SA algorithm consists of performing a sequence of stochastic configuration sampling of the search space  
using an 
acceptance criterion of the new configuration based on the Boltzmann probability distribution at temperature $T$ while progressively reducing $T$. 
A system in thermal equilibrium at temperature $T$ can be found in a state with a cost function (the Hilbert-Schmidt distance in our case) $D$ with a probability proportional to $\exp (-D / T)$. 
At low temperatures, there is a small probability that the system will be in a high-cost-function state. 
This plays a crucial role in the procedure because a possible increase in the cost function allows the system to escape from local minima and potentially find the global minimum. 
According to our experience, a good choice for $T$ is in the range $[0,0.1]$ (atomic units will be used throughout) depending on the system and target $p$-body matrix, 
allowing to  avoid possible local minima and, at the same time,  keeping the number of iterations manageable.
The  decreasing rate of $T$ for each step $i$ is chosen so that  $T^{(i+1)}= \delta T\,T^{(i)}$  with  $\delta T= 0.995$, while $\theta_{\mathrm{max}}^{(i+1)}=\delta \theta_{\mathrm{max}}\, \theta_{\mathrm{max}}^{(i)}$, with
an  initial value $\theta_{\mathrm{max}}= 0.5$ and $\delta \theta_{\mathrm{max}} = 0.999$
when the new configuration is not accepted or $\delta \theta_{\mathrm{max}} = 1.0025$ otherwise. 
This strategy effectively 
decreases $\theta_\mathrm{max}$ slowly while still allowing the system to escape local minima. 

The variational 2-body-reduced-density-matrix (v2RDM) methods relying on semidefinite programming \cite{Mazziotti.book.2007,Coleman.book.2000,doi:10.1021/cr2000493,DePrince.2024,Nakata.jcp.2001, Mazziotti.prl.2004, Gidofalvi.jcp.2008,Mazziotti.prl.2011, Fosso.jctc.2016}
within a SU(2) pairing algebra, or doubly occupied configuration interaction (DOCI) framework, \cite{DePrince.2024,Weinhold.JCP.1967,Weinhold.JCP.1967.II,Poelmans.JCTC.2015,Alcoba.JCP.2018,Rubio-Garcia.JCTC.2018,Alcoba.JCP.2018.149,Rubio-Garcia.JCP.2019}}
with different $N$-representability conditions were utilized to obtain (non $N$-representable) approximated 2-body reduced density matrices of the ground state of the reduced BCS and Heisenberg XXZ spin models. 
The resulting matrices were then used as targets for the ADAPT-VQA to  analyze the numerical effectiveness of the method 
by comparing the v2RDM solutions.
The v2RDM method provides a lower-bound energy to the exact ground state, and it has been utilized to validate different sets of $N$-representability conditions.\cite{DePrince.2024}
The constraints typically utilized in v2RDM methods are 
not restrictive enough to ensure the exactness of the resulting
2-body matrices, hereafter denoted as $^2\rho_{\mathrm{v2RDM}}$.
In our case,  we obtain the $^2\rho_{\mathrm{v2RDM}}$
applying 2-positivity and partial 3-positivity $\mathrm{T1}$ and $\mathrm{T2}^\prime$ conditions
\cite{Nakata.jcp.2001, Mazziotti.pra.2002, Zhao.jcp.2004, Mazziotti.pra.2005}
reformulated using the SU(2) pairing algebra \cite{Weinhold.JCP.1967,Weinhold.JCP.1967.II,Poelmans.JCTC.2015,Alcoba.JCP.2018,Rubio-Garcia.JCTC.2018,Alcoba.JCP.2018.149,Rubio-Garcia.JCP.2019},
which are two of the most restrictive sets of constraints currently available for practical calculations,
with the latter leading to a more stringent combined set and therefore significantly improving the accuracy of the 2-positivity conditions \cite{Mazziotti.PRL.2012}.
These two sets of constraints, hereafter denoted as 2-POS and (2,3)-POS,  are
complemented  with hermiticity and normalization conditions 
on $^2\rho_{\mathrm{v2RDM}}$,
and  with the corresponding contraction and consistency relations for each set of conditions.\cite{Poelmans.JCTC.2015,Alcoba.JCP.2018,Rubio-Garcia.JCTC.2018,Rubio-Garcia.JCP.2019}



\par\bigskip
{\it Linear H$_4$ molecule:} Within the Born-Oppenheimer approximation, the non-relativistic Hamiltonian of an  $N$-electron molecular system can be written within the second quantization formalism as
\cite{Surjan.book.1989}

\begin{equation}
H = \sum_{ij}\;\langle i | h |j\rangle\;a^{\dag}_{i}a_{j} +
\frac{1}{4}\;\sum_{ijkl}\; \langle ij| v |kl\rangle\;a^{\dag}_{i}a^{\dag}_{j}a_{l}a_{k}\,,
\label{eq:Hgen}
\end{equation}
where $\langle i| h |j\rangle$ and $\langle ij| v |kl\rangle $ are the standard
one- and two-electron antisymmetrized integrals, respectively, and  $a^\dagger_i$ and $a_j$ are the
 fermion creation and annihilation operators acting
on an orthonormal $2K$-dimensional single-particle 
spin-orbital basis built from $K$ spatial orbitals. 
For the case of the linear H$_4$ molecule at
  0.75~{\AA} inter-atomic separation,  the exact (within the basis set) singlet ground- and first-excited full configuration interaction solutions are used to obtain the corresponding
exact 1-body reduced density matrices, denoted as $^1\rho_{\mathrm{exact}}$, and the exact ground-state 2-body reduced density matrix, denoted as $^2\rho_{\mathrm{exact}}$, respectively. 
The resulting ground state  resembles a closed-shell singlet with a single leading determinant coefficient of  $\sim$0.98, while the first excited stated is of multi-reference character, closely resembling a spin-adapted singlet
with two dominating determinants with the largest coefficient of $\sim$0.70.
The eigenvalues of the exact ground state $1$-RDM lie in the interval [0.0048,0.9936], while those for the excited state are in the interval [0.0015, 0.9985]. On the other hand, the eigenvalues of the exact ground-state $2$-RDM were found in the interval [0.0000,2.0148].

\begin{figure*}[h]
\centering
\includegraphics[width=0.9\linewidth]{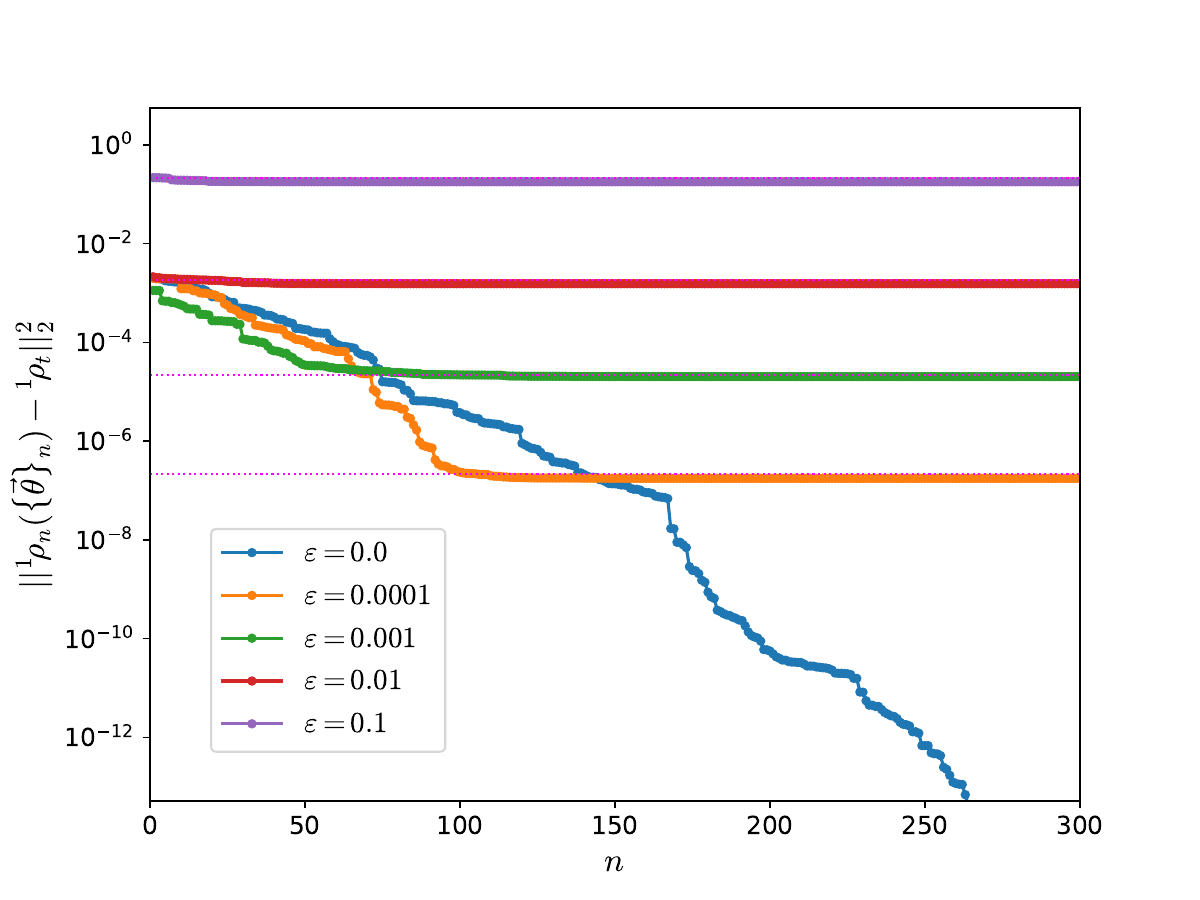}
\caption{Distance between the 1-body reduced density matrix $^1\rho(\{\vec{\theta}\})$ and a fixed noisy target   
$^1\rho_t$ as a function of iteration number $n$ for the linear H$_4$ molecule. 
The target is constructed by adding random noise of strength  $\varepsilon$ to the exact ground-state $^1\rho_{\mathrm{exact}}$.
Distances between the exact (unperturbed) ground-state $1$-body reduced density matrix and the fixed targets  $^1\rho_t$ are shown as upper-bound references with pink dotted lines. The initial $\rho_0$ is constructed from the Hartree-Fock ground state.
\label{fig1}}
\end{figure*}

Our first test was to monitor the evolution of the distance $D$ 
 between the (physical) reduced state of  $\rho(\{\vec{\theta}\})$ on a $1$-body subsystem, represented by a $1$-body reduced density matrix $^1\rho(\{\vec{\theta}\})\,\equiv\, N\Tr_{N-1}[\rho(\{\vec{\theta}\})]
$ and a fixed target one, represented by  
$^1\rho_t$, as a function of iteration number $n$. We build a noisy target 1-body matrix from $^1\rho_{\mathrm{exact}}$ as   $\;^1\rho_t = \;^1\rho_{\mathrm{exact}} + \varepsilon  \;R$, with $R$ a matrix of random numbers taken from a uniform probability distribution in $[-1,1]$, and $\varepsilon$ the strength of perturbation
in the interval $[0,0.1]$, which must be compared with the extreme eigenvalues of the (unperturbed) exact 1-RDM. The ADAPT-VQA is expected to evolve 
the reduced state of $\rho_0$, initialized as the Hartree-Fock ground state, on a 1-body subsystem
towards the noisy  $^1\rho_{t}$. Since the noise in the target 
breaks the $N$-representability of the exact 1-RDM, the algorithm should only approach $^1\rho_{n}$ to the target up to a certain limit, as seen in Fig.~\ref{fig1}. Larger noise strengths  $\varepsilon$ lead to convergence of the algorithm to larger $D$, while a noiseless target allows $^1\rho_{n}$  to approach $^1\rho_{t}$ to 10$^{-12}$ in $\sim$250 iterations.

\begin{figure*}[h]
\centering
\includegraphics[width=0.9\linewidth]{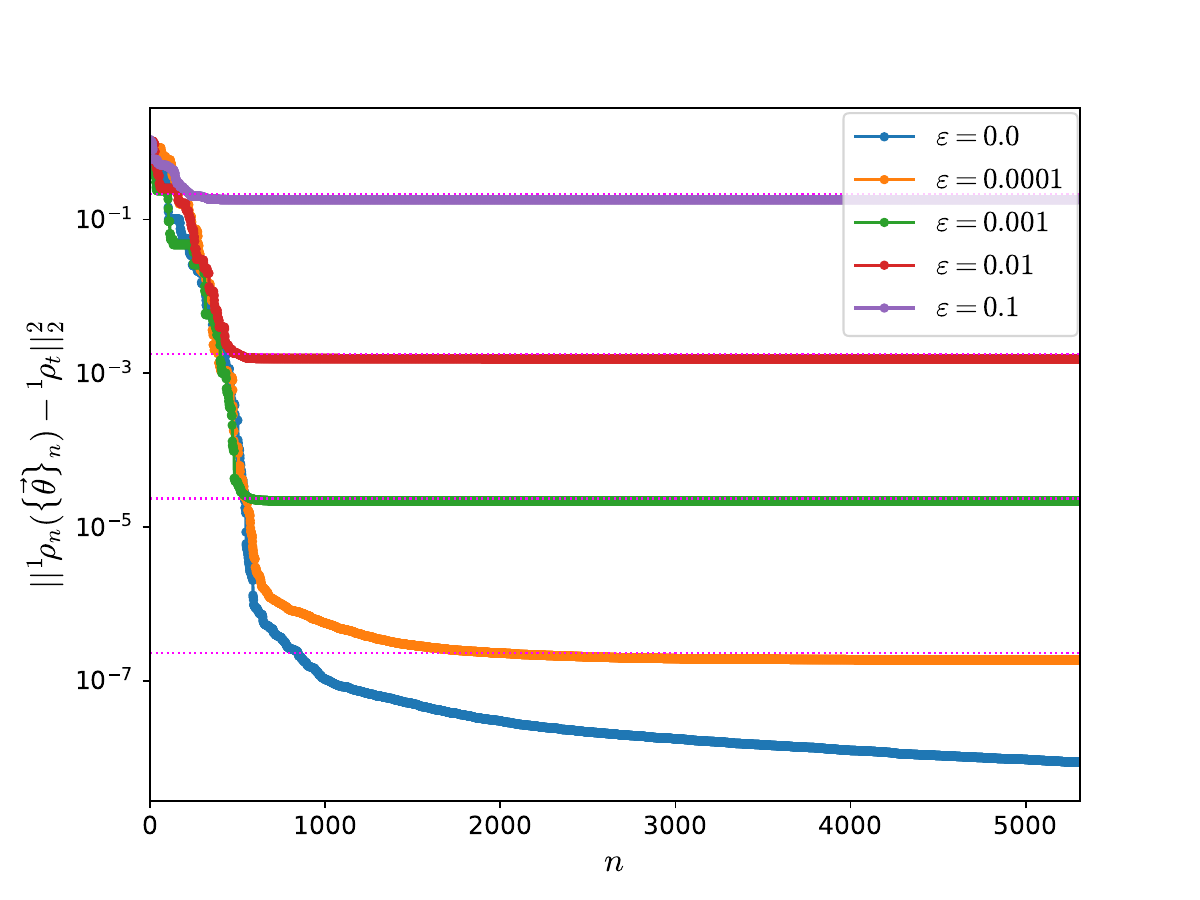}
\caption{Distance between the 1-body reduced density matrix $^1\rho(\{\vec{\theta}\})$ and a fixed noisy target   
$^1\rho_t$ as a function of iteration number $n$ for the linear H$_4$ molecule. 
The target is constructed by adding random noise of strength  $\varepsilon$ to the exact first excited state $^1\rho_{\mathrm{exact}}$.
Distances between the exact (unperturbed) first-excited-state $1$-body reduced density matrix and the fixed targets  $^1\rho_t$ are shown as upper-bound references with pink dotted lines. The initial $\rho_0$ is constructed from the Hartree-Fock ground state. \label{fig2}
}
\end{figure*}

As a second test, we repeated the previous numerical experiment but using the 1-body reduced density matrix corresponding to the exact first-excited state, also denoted as $^1\rho_{\mathrm{exact}}$, while keeping $\rho_0$  as the Hartree-Fock ground-state. 
Perturbation strength $\epsilon$ was also considered the interval $[0,0.1]$, which must be compared against the extreme eigenvalues of the (unperturbed) exact 1-RDM.
Here, one expects a slower convergence, as shown in Fig.~\ref{fig2}, but nonetheless, the ADAPT-VQA performs as expected. Approximately  6000 simulated-annealing iterations are needed to reach a distance of 10$^{-8}$ to the noiseless target. 

Importantly,
initializing the ADAPT-VQA with a correlated state, the algorithm converges to the same solution obtained with the Hartree-Fock state for both cases, the ground and excited target matrices (see Supporting Information for details).

\begin{figure*}[h]
\centering
\includegraphics[width=0.9\linewidth]{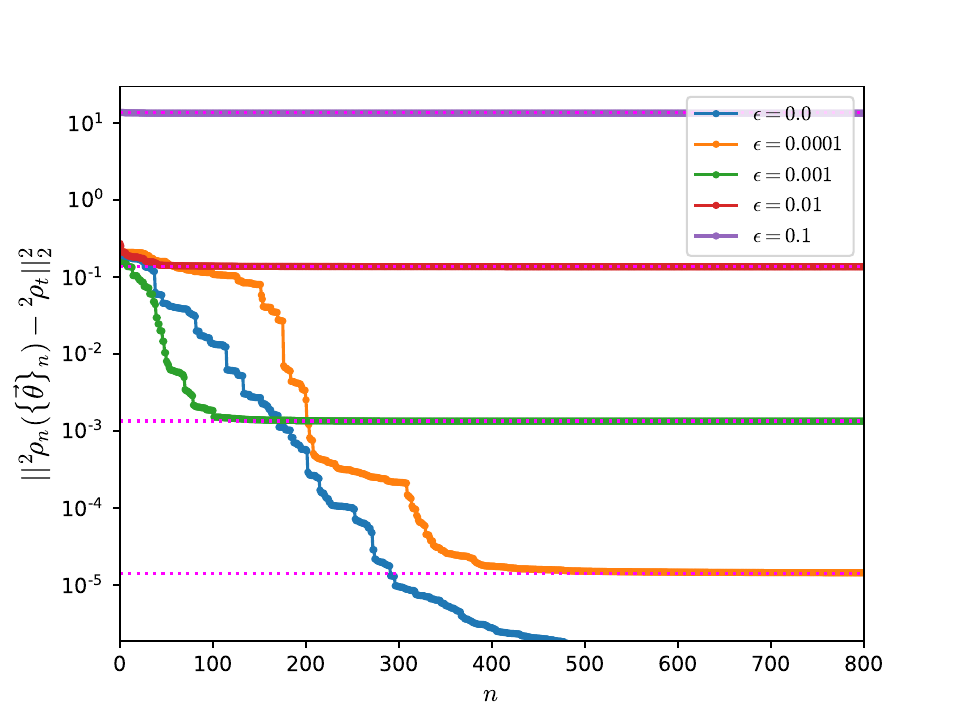}
\caption{Distance between the 2-body reduced density matrix $^2\rho(\{\vec{\theta}\})$ and a fixed noisy target   
$^2\rho_t$ as a function of iteration number $n$ for the linear H$_4$ molecule. 
The target is constructed by adding random noise of strength  $\varepsilon$ to the exact  ground-state $^2\rho_{\mathrm{exact}}$.
Distances between the exact (unperturbed) ground-state $2$-body reduced density matrix and the fixed targets  $^2\rho_t$ are shown as upper-bound references with pink dotted lines. The initial $\rho_0$ is constructed from the Hartree-Fock ground state. 
\label{fig3}
}
\end{figure*}

Next, we examine the behavior of our ADAPT-VQA for $2$-body reduced density matrices. 
To this end, we analyze the convergence of $^2\rho_{n}$ towards the  $^2\rho_{\mathrm{exact}}$ with noise added as in the two cases before, which must be compared against the extreme eigenvalues of the (unperturbed) exact 2-RDM. Again, $\rho_0$ is initialized as the Hartree-Fock ground state. The stochastic process allows
$^2\rho_{n}$ to approach the target 2-RDM progressively up to a certain limit determined by the level of noise in the target, as shown in Fig.~\ref{fig3}. The noiseless case reaches a value of $D\sim$10$^{-5}$ in $\sim$300 iterations. Here it should be reminded that the dimension of the 2-RDM problem is larger than that corresponding to the 1-RDM, and thus a larger number of simulated-annealing steps are expected to reach similar convergence. The three examples reported for the H$_4$ molecule show that the algorithm works as expected for electronic systems.

\par\bigskip

{\it Reduced BCS model:} 
The reduced BCS or constant pairing model is employed in nuclear \cite{Dean2003} and condensed matter physics \cite{Delft.prl.1966,Sierra.PRL.2000} to model the superfluid and superconducting properties of finite and extended systems, respectively. In this work, we use the reduced BCS Hamiltonian in the form
\cite{Richardson.physLett.1963,Rubio-Garcia.JCTC.2018}
\begin{equation}
  H = \sum_{i}\;\epsilon_{i}n_{i} -\;G\;\sum_{ij}\; b^{\dag}_{i}b_{j}\,,
\label{eq:Hdoci}
\end{equation}
where $G$ is the strength of the infinite-range pairing interaction, which may be repulsive ($G < 0$) or attractive ($G > 0$), and $\epsilon_{i}$ are single-particle energies. 
The number operator is
$n_i=\frac{1}{2}(a_i^{\dagger }a_i+a_{\bar{i}}^{\dagger}a_{\bar{i}})$, and the
pair-creation (annihilation) operators are
$b^{\dagger}_{i}=\left(b_{i}\right)^{\dagger}=a_{i}^{\dagger}a_{\bar{i}}^{\dagger}$.
We will only
consider half-filled states with the number of pairs
$M=K/2$ (where $K$ is the total number of single-particle levels) and
equally-spaced single-particle energies $\varepsilon_{i} =i/K$, with 
$i=1,2,\ldots,K$. 
The $(i,\bar{i})$ pair defines the pairing scheme, which involves
two particles with opposite conjugate
quantum numbers (i.e. spin or momentum) in doubly degenerate single-particle levels.
The Hamiltonian (\ref{eq:Hdoci}) is based on the $SU(2)$ pairing algebra with generators $b_i^{\dagger}$ and $(2n_i-1)/2$, and hard-core boson relations \cite{Matsubara.PTP.1956,Matsubara.PTP.1957}

\begin{equation} [b_i,b_j^{\dagger}] = \delta_{ij}(1-2n_i),\quad
  (b_i^{\dagger})^2=0.
\end{equation}

\begin{figure*}[h]
\centering
\includegraphics[width=0.9\linewidth]{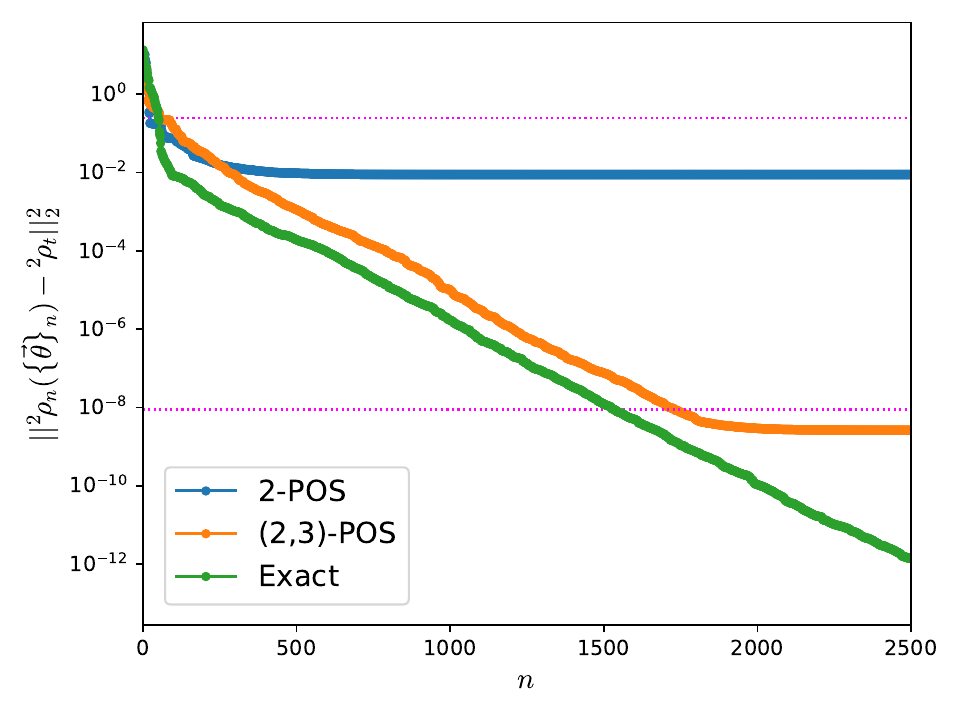}
\caption{Distance between the 2-body reduced density matrix $^2\rho(\{\vec{\theta}\})$ in the reduced BCS model (see text) and three different targets obtained from the v2RDM methods with  2-POS and (2,3)-POS conditions, and the exact ground-state solution. Distances between the exact ground-state $2$-body reduced density matrix and the fixed targets  $^2\rho_t$ are shown as upper-bound references with pink dotted lines.} The initial  $\rho_0$ is constructed from the mean-field $G=0$  pairing model ground state.\label{5}
\end{figure*}

This model can be solved exactly, \cite{Richardson.physLett.1963}
predicting a gapless phase (metallic) and a finite gap phase (superconducting). 
The critical value of the strength parameter $G$ that separates these phases, $G_c$, is obtained from the gap equation in the zero-gap limit and the chemical potential $\mu =(\varepsilon_M + \varepsilon_{M+1})/2$,
\begin{equation}
G_c = \left[ \sum_i \frac{1}{ \mid \varepsilon_i - \mu \mid}  \right]^{-1}\,.
\label{29}
\end{equation}
\noindent
For the reduced BCS model, we first explore the convergence of $^2\rho_n$ towards a target 2-body matrix calculated using three different approaches: The v2RDM methods enforcing 2-POS and (2,3)-POS conditions, and the 2-RDM corresponding to the exact ground-state. These simulations employ an interaction strength $G=1$ for a system of $K = 4$ single-particle levels at half-filling, and the initial $\rho_0$ is initialized from a simple independent-particle-model solution, the non-interacting $G=0$ pairing model ground state. 
In this case, the BCS model with critical value, $G_c =  0.1875$, is a sensible example to assess the extent of pairing correlations. 
Hence, we have considered pairing strengths around this value.
As in the previous cases, our stochastic  ADAPT-VQA evolves $^2\rho_n$ towards the target matrix, Fig.~\ref{5}. In the case of the exact ground-state target, the distance reduces to 10$^{-12}$ in about 2500 iterations, while with targets obtained from the v2RDM methods, the distance plateaus to small, yet numerically significant values:  about 10$^{-8}$ for the  (2,3)-POS and 10$^{-2}$ for the  2-POS. This illustrates the potential of our approach as a tool to quantify the quality of the approximated reduced density matrices provided by different approaches.

\begin{figure*}[h]
\centering
\includegraphics[width=0.9\linewidth]{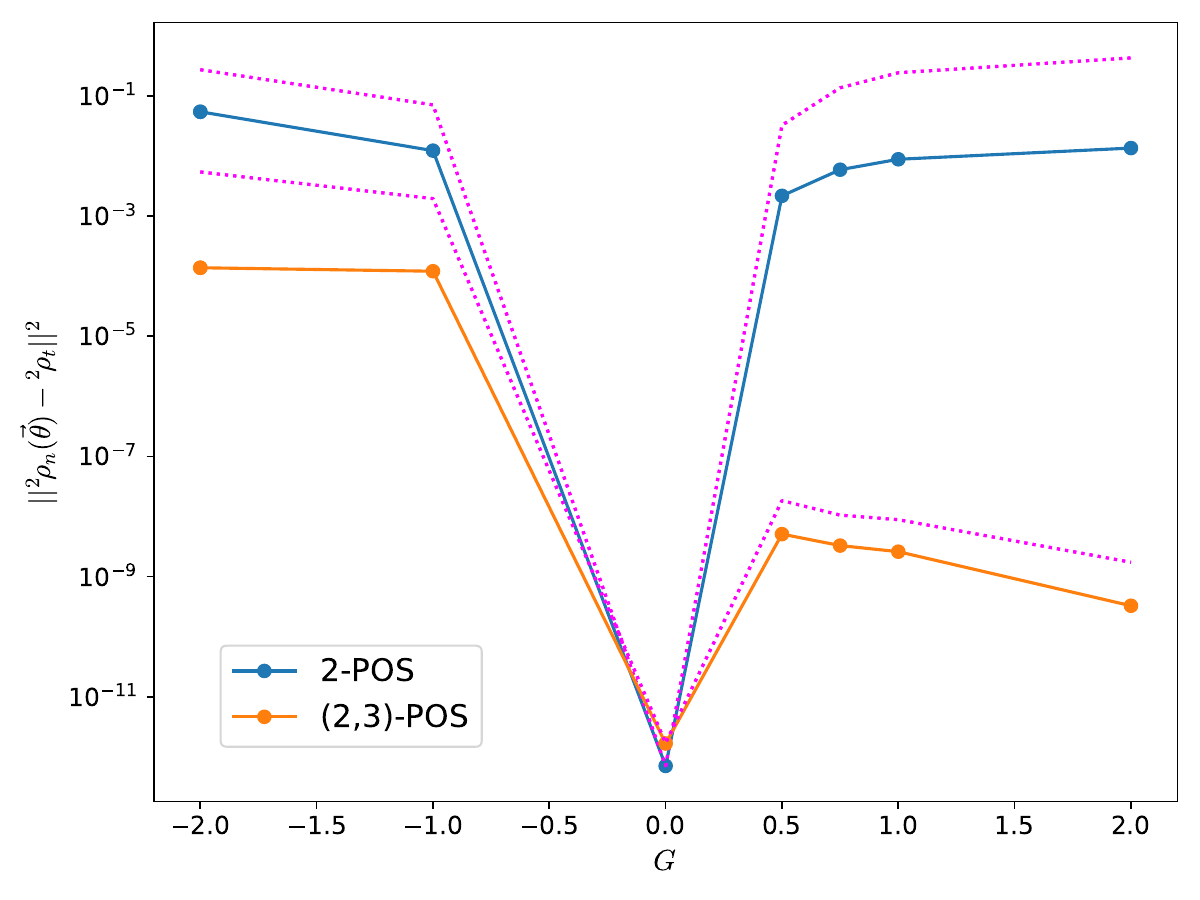}
\caption{
Distance between the fully evolved 2-body reduced density matrix obtained with the ADAPT-VQA and a fixed target calculated from the  2-POS and (2,3)-POS v2RDM methods for the reduced BCS model. 
Distances between the exact ground-state $2$-body reduced density matrix and the fixed targets  $^2\rho_t$ are shown as upper-bound references with pink dotted lines. The initial  $\rho_0$ is constructed from the mean-field $G=0$  pairing model ground state. \label{6}}
\end{figure*}

As a second test for the reduced BCS model, we now analyze the converged distance the ADAPT-VQA gives for interaction strengths $G$ in the range of [-2,2]. Here we include $G=0$ (non-interacting) and $G=0.5, 0.75, \pm1,\pm2$ and plot in Fig.~\ref{6} the final converged distance. The v2RDM 2-RDM with  2-POS and (2,3)-POS conditions are targets, starting from the non-interacting ($G = 0$) state for the different G values.
The ADAPT-VQA can evolve the non-interacting 2-RDM towards the approximate variational solutions for different interaction strengths. The 2-POS target shows a rather symmetric behavior around $G=0$, with a large converged distance in the vicinity of 10$^{-1}$ for $G\ne0$. On the other hand, the (2,3)-POS target shows an asymmetric behavior about $G=0$, with larger distances of $\sim$10$^{-4}$ for $G<0$ and significantly reduced converged distances for $G>0$, $\sim$10$^{-9}$. 
Figure \ref{6} shows that both 2-POS and (2,3)-POS target 2-RDMs corresponding to $G=0$ are numerically $N$-representable, in agreement with the fact that for $G=0$ the set of 2-POS (and consequently (2,3)-POS) $N$-representability conditions together with the hermiticity, contraction, and consistency conditions become necessary and sufficient,
while including correlation ($G\ne0$) brings the 2-POS and (2,3)-POS targets farther from an  $N$-representable 2-RDM.  This highlights the potential use of the ADAPT-VQA to establish  a numerical measure 
of $N$-representability.

\par\bigskip

\begin{figure*}[h]
\centering
\includegraphics[width=0.9\linewidth]{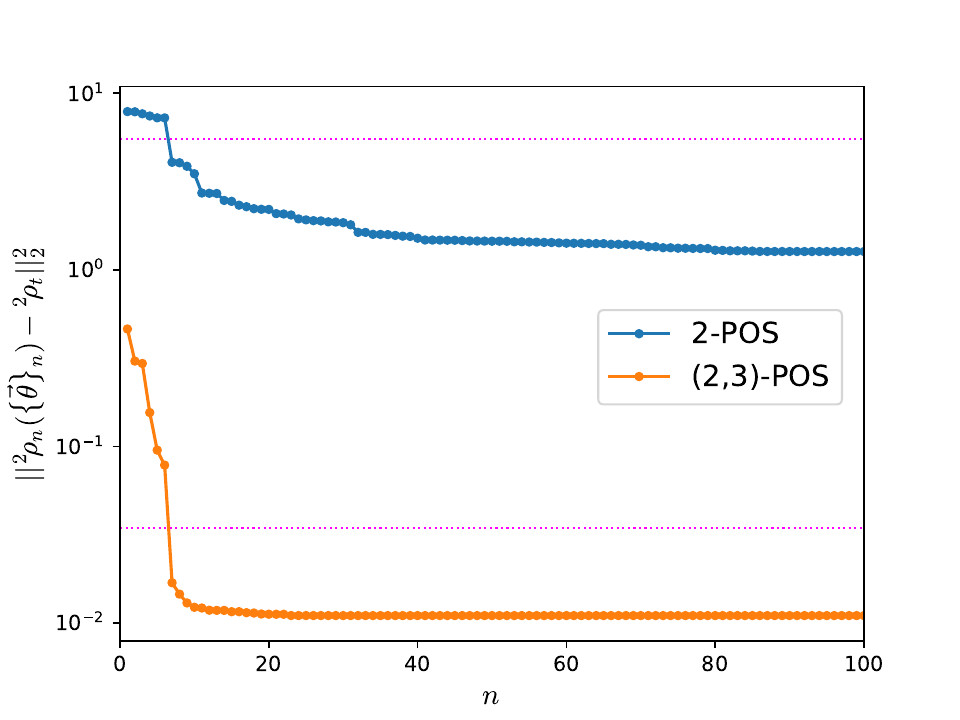}
\caption{
Distance between the 2-body reduced density matrix $^2\rho(\{\vec{\theta}\})$ in the  XXZ spin model with anisotropy $\Delta=2$  (see text) and two target matrices obtained from the v2RDM methods with  2-POS and (2,3)-POS conditions.
Distances between the exact ground-state $2$-body reduced density matrix and the fixed targets  $^2\rho_t$ are shown as upper-bound references with pink dotted lines. The initial  $\rho_0$ is chosen as the exact ground state of the $\Delta=\infty$ XXZ model.\label{7}
}
\end{figure*}

{\it Heisenberg XXZ model:} 
The Heisenberg Hamiltonian can be used to model the low-excitation physics of 
a number of problems, such as quantum magnetism and  Mott phase transitions.
\cite{auerbach2012}
The Hamiltonian with anisotropy of  XXZ type for a chain
of $K$ spin sites  with open boundary conditions with first-neighbor interactions can be cast as
\begin{align}
\label{eq14}
  H
    &= \sum_i\left[\frac{1}{2}(S_i^{+}S_{i+1}^{-} + S_i^{-}S_{i+1}^{+}) + \Delta S_i^zS_{i+1}^z\right]
\end{align}
where $S_i^{\pm}$ and $S_i^{z}$ are the fermionic spin-ladder
and spin projection operators acting on site $i$, and   
the parameter
$\Delta$ sets the anisotropy of the model.  
We consider a finite chain of 4 sites at $S^z_{tot}=0$ with open boundary conditions.  This model has a rich phase diagram as a function of $\Delta$. For $-1<\Delta<1$, the system is a
critical antiferromagnet with gapless excitations. 
For $|\Delta| > 1$,
the system shows a finite energy gap, and  it is ferromagnetic for $\Delta<-1$
and antiferromagnetic for $\Delta > 1$. 
Although the Hamiltonian in Eq.~(\ref{eq14}) can be straightforwardly implemented in a quantum computer, we opt here to map it to a 
 hard-core bosons
representation as introduced by Holstein and Primakoff.\cite{Holstein.PR.1940}
This choice allows us to directly assess the quality of variational 2-RDMs obtained using an $SU(2)$ pairing algebra formalism.\cite{Rubio-Garcia.JCP.2019,Massaccesi.JSM.2021}
Using this
representation, the spin operators are
\begin{equation}
  S_i^+=b_i^{\dagger}=(S_i^-)^\dagger,\quad S_i^z=b_i^{\dagger}b_i-\frac{1}{2}=n_i-\frac{1}{2}
\end{equation}
and the total number of hard-core bosons relates to the spin $z$ projection  as $\sum_i n_i = M=S_{tot}^z + K/2$, making  half-filling
($M/K=1/2$) correspond to $S_{tot}^z=0$. 
Using this transformation, the Hamiltonian reads
\begin{align}
  H &=\sum_{i}\left[\frac{1}{2}(b_i^{\dagger}b_{i+1} + b_{i+1}^{\dagger}b_i) + \Delta \left(n_i-\frac{1}{2}\right) \left(n_{i+1}-\frac{1}{2}\right)\right].
      \label{eq:HXXZ}
\end{align}
Similar to the case of the reduced BCS model, in this case, we first analyze the convergence of a 2-RDM evolved with our ADAPT-VQA towards fixed targets. The targets are chosen as the 2-POS and (2,3)-POS v2RDM solutions as before, while the initial state $\rho_0$ is initialized as the $\Delta=\infty$ exact ground state. The results for $\Delta=2$ are summarized in Fig.~\ref{7}. In this case, convergence is achieved quicker than in all previous numerical tests. However,  it becomes evident from Fig.~\ref{7} that the 2-POS and (2,3)-POS v2RDM reduced density matrices are somewhat poor approximations in terms of the quality of the matrices, giving converged distances of $\sim 1$ and $\sim$10$^{-2}$, respectively. Other authors have observed this behavior for the XXZ model using semi-definite relaxation variational RDM techniques.
\cite{Rubio-Garcia.JCP.2019,Massaccesi.JSM.2021,Barthel.PRL.2012,Haim.2020}

\begin{figure*}[h]
\centering
\includegraphics[width=0.9\linewidth]{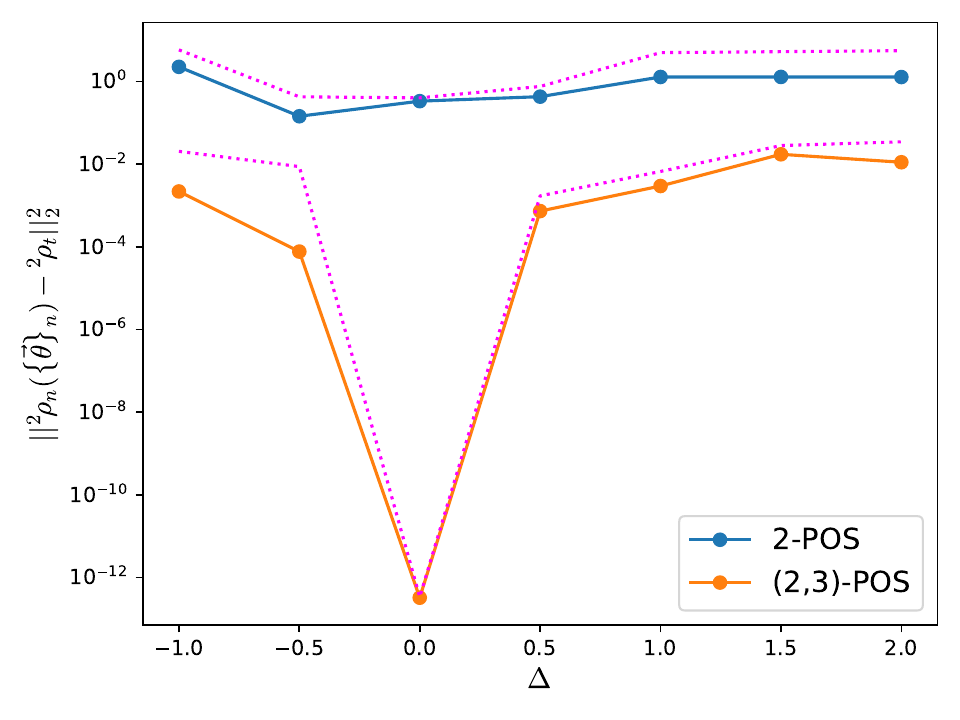}
\caption{
\label{8}
Distance between the fully evolved 2-body reduced density matrices obtained with the ADAPT-VQA and fixed targets calculated from the v2RDM methods using 2-POS and (2,3)-POS conditions of the XXZ spin model as a function of the anisotropy $\Delta$. The targets are the 2-body reduced density matrices computed variationally with 
2-positivity (2-POS) and partial 3-positivity ((2,3)-POS)
conditions.
Distances between the exact ground-state $2$-body reduced density matrix and the fixed targets  $^2\rho_t$ are shown as upper-bound references with pink dotted lines.
}
\end{figure*}

As a last test, we compare the fully converged 2-body reduced density matrix distance given by our ADAPT-VQA for different anisotropy strengths $\Delta$. We utilize, as before, the two target variational approximate 2-body reduced density matrices obtained from the v2RDM methods with  2-POS and (2,3)-POS conditions. Fig.~\ref{8} shows the results for $-1<\Delta<2$. For the case of $\Delta=0$, which corresponds to the isotropic model, the  ADAPT-VQA can evolve the initial 2-RDM to a numerically negligible distance from the (2,3)-POS v2RDM matrix. This is not the case for other anisotropy values and for the 2-POS v2RDM matrix. There, the ADAPT-VQA evolves the initial 2-RDM to distances of $\sim 1$ and 10$^{-2}$ to 10$^{-4}$ for the 2-POS and (2,3)-POS cases, respectively.

\section{Summary and Outlook}\label{sec13}

We have introduced a hybrid ADAPT-VQA that evolves an initial $N$-body density matrix to successively approach the reduced state of the density matrix on a $p$-body subsystem, represented by a $p$-RDM, to a target $p$-body matrix, alleged $p$-RDM.
The algorithm makes use of the  ADAPT methodology to progressively construct a quantum circuit, represented by 
 a set of vector
parameters and operators to generate an $N$-body density matrix. 
The optimization  
relies on a stochastic simulated annealing procedure, thus avoiding potential barren plateaus that are prone to gradient-descent methods. 
Importantly, the proposed ADAPT-VQA is independent of any underlying Hamiltonian and can be used to determine the quality and correct an alleged $p$-body reduced density matrix. It is robust under statistical noise. 

We have challenged the robustness of the proposed ADAPT-VQA with 1- and 2-RDMs using the linear H$_4$ quantum chemistry electronic Hamiltonian, the reduced BCS model with constant pairing, and the Heisenberg XXZ spin model. 
Using these proof-of-concept cases, we illustrate the applicability of the algorithm considering starting points and different targets. 
We found that the ADAPT-VQA behaves as expected for pure 1- and 2-RDMs, evolving the initial matrices towards the target.  
The extension of the algorithm to evolve initial ensemble states and their corresponding $p$-RDMs, and hence to tackle the ensemble N-representability problem, will be reported elsewhere.
Overall, the proposed ADAPT-VQA provides a valuable addition to the arsenal of electronic structure quantum algorithms.

\begin{suppinfo}
Comparison of the convergence behavior of the ADAPT-VQA for different initial states.
\end{suppinfo}

\begin{acknowledgement}
GEM, OBO, PC, JIM, and DRA acknowledge the financial support from the Consejo Nacional de Investigaciones Cient\'{\i}ficas y T\'ecnicas (grant No. PIP KE3 11220200100467CO and PIP KE3 11220210100821CO). GEM, OBO, PC, and DRA acknowledge support from the Universidad de Buenos Aires (grant No. 20020190100214BA and 20020220100069BA)) and the Agencia Nacional de Promoci\'on Cient\'{\i}fica y Tecnol\'ogica (grant No. PICT-201-0381). JEP acknowledges support from the U.S. Department of Energy, Office of Basic Energy Sciences, as part of the Computational Chemical Sciences Program under Award No. DE-SC0005027. GES is a Welch Foundation Chair (C-0036), and his work was supported by the U.S. Department of Energy under Award No. DE-SC0019374. 
\end{acknowledgement}

\providecommand{\latin}[1]{#1}
\makeatletter
\providecommand{\doi}
  {\begingroup\let\do\@makeother\dospecials
  \catcode`\{=1 \catcode`\}=2 \doi@aux}
\providecommand{\doi@aux}[1]{\endgroup\texttt{#1}}
\makeatother
\providecommand*\mcitethebibliography{\thebibliography}
\csname @ifundefined\endcsname{endmcitethebibliography}  {\let\endmcitethebibliography\endthebibliography}{}

\end{document}